# Evidence of Spin-Valley Coupling in Dirac Material $BaMnBi_2$ Probed by Quantum Hall Effect and Nonlinear Hall Effect

Subin Mali[1], Yingdong Guan[1], Lujin Min[1,2], David Graf[3], and Zhiqiang Mao[1,2*]

[1]Department of Physics, Pennsylvania State University, University Park, PA 16802, USA

[2]Department of Materials Science and Engineering, The Pennsylvania State University, University Park, PA 16802, USA

[3]National High Magnetic Field Lab, Tallahassee, Florida 32310, USA

## ABSTRACT

Valleytronics is a rapidly advancing field that explores the use of the valley degree of freedom in electronic systems to encode and process information. It relies on electronic states with spin-valley locking, first predicted and observed in monolayer transition metal dichalcogenides like $MoS_2$. However, very few bulk materials have been reported to host spin-valley locked electronic states. In this work, we present experimental evidence for a predicted, unique spin-valley locked electronic state generated by the Bi zig-zag chains in the layered compound $BaMnBi_2$. We observed remarkable quantum transport properties in this material, including stacked quantum Hall effect (QHE) and nonlinear Hall effect (NLHE). From the analysis of the QHE, we identified a spin-valley degeneracy of 4, while the NLHE provides supporting evidence for the anticipated valley-contrasted Berry curvature—a typical signature of a spin-valley locked state. This spin-valley locked state contrasts with that observed in the sister compound $BaMnSb_2$, where the degeneracy is 2. This difference arises from significant variations in their orthorhombic structures and spin-orbital coupling. These findings not only set up a new platform for exploring coupled spin-valley physics in bulk materials but also underscores its potential for valleytronic device applications.

*Email: zim1@psu.edu

## I. INTRODUCTION

Non-centrosymmetric materials with strong spin-orbit coupling (SOC) often display electronic states characterized by spin splitting of energy bands. When the valley degree of freedom is incorporated into these materials, the spin and valley degrees of freedom become coupled (also termed spin-valley locking), creating unique electronic states. This coupling allows for the control and manipulation of the valley degree of freedom [1] using electric fields [2], magnetic fields [3], or light [4]. These unconventional electronic states hold great promise for a wide range of potential applications, including valleytronic logic gates and valley-based memory devices [5,6].

The spin–valley locked state was first predicted in $MoS_2$ monolayers [1] and later demonstrated experimentally through optical helicity-controlled valley polarization [7], optically driven valley polarization [8], and spin–pseudospin spectroscopy [9,10]. The inversion symmetry is broken in monolayers, and the conduction and valence band edges are located at the corners (K points) of the 2D hexagonal Brillouin zone. The combination of strong SOC and inversion symmetry breaking in these materials enables spin splitting at K and K' valleys with opposite spin orientation, leading to the valley dependent spin polarization, i.e. spin-valley locking. Spin-valley coupling in TMDCs has given rise to various interesting phenomena, such as valley-dependent circular dichroic photoluminescence [8,11,12], nonreciprocal charge transport [13], photo-induced charge Hall effect [1], valley Hall effect [14], and exciton Hall effect [15] in the absence of an external magnetic field. In the bulk form of TMDCs, there have been two reported examples that demonstrate spin-valley locking, namely 3R-$MoS_2$ [16] and 2H-$NbSe_2$ [17]. Besides TMDCs, layered polar Dirac material $BaMnSb_2$ has been identified as the only known bulk material exhibiting spin-valley locking beyond TMDCs [18]. In $BaMnSb_2$, the Sb zig-zag chain layers break in-plane inversion symmetry and combined with the strong SOC from the Sb atoms, two Dirac valleys with valley-contrasting out-of-plane spin polarizations are formed around the X point. This was revealed through first-principles calculations, angle-resolved photoemission spectroscopy (ARPES), and quantum Hall measurements [18–20].

To further understand coupled spin and valley physics and explore valleytronic and optoelectronic applications of spin-valley locked states, it is crucial to find various spin-valley locked states in more bulk materials. Despite the availability of numerous materials with inversion symmetry breaking and strong SOC, the limited number of known candidates underscores the challenge in identifying new materials with this unique property. Recently, Kondo et al. [21] used crystallographic analysis, first-principles calculations, and quantum oscillation measurements to investigate $BaMnBi_2$, a sister compound of $BaMnSb_2$. Their work predicted the existence of multiple pairs of Dirac valleys featuring valley dependent spin polarization and reported that, similar to $BaMnSb_2$, $BaMnBi_2$ also exhibits a polar orthorhombic structure with the space group *Imm2* (No. 44) [21]. Its orthorhombic distortion leads Bi to form zigzag chain layer which hosts Dirac fermions and is sandwiched between the insulating $Ba^{2+}$-$Mn^{2+}$-$Bi^{3-}$-$Ba^{2+}$ slabs (Fig. 1a). However, $BaMnBi_2$ exhibits a slightly contracted *c*-axis, but significantly reduced orthorhombicity compared to $BaMnSb_2$ [21]. These structure characteristics, combined with stronger SOC of Bi, lead $BaMnBi_2$ to possess a spin-valley locked Dirac state distinct from that of $BaMnSb_2$. $BaMnSb_2$ exhibits a single pair of valleys with spin-valley locking near the X-point on the zone boundary (Fig. 1b, left panel). In contrast, the spin-valley locked state in $BaMnBi_2$ is reported to possess four and two spin-polarized valleys near the X and Y points, respectively, as well as four additional valleys near the midpoint of the ΓM line (Fig. 1b, right panel) [21]. DFT calculations and Shubnikov–de Haas (SdH) oscillations reported in Ref. [21] provide strong evidence for spin-split bands in $BaMnBi_2$. While these results represent a significant advance in understanding its electronic structure, direct experimental demonstration of topological quantum transport arising from this unique spin–valley–locked state, as well as identification of the spin–valley degeneracy, has yet to be achieved.

Here we report on quantum transport properties of $BaMnBi_2$. Through systematic measurements conducted on this material, we not only observed a stacked quantum Hall effect (QHE) but also detected an intrinsic nonlinear Hall effect (NLHE) [22]. The QHE analysis revealed a spin-valley degeneracy of 4, indicating the presence of two pairs of spin-valley locked Dirac cones near the Fermi level in $BaMnBi_2$, contrasted with the spin-valley locked state with a degeneracy of 2 observed in $BaMnSb_2$. Moreover, the observation of intrinsic NLHE in bulk $BaMnBi_2$ single crystals suggests the existence of a Berry curvature dipole [23,24] in its band structure, providing strong supporting evidence for the expected valley-contrasted Berry curvature in a spin-valley locked state. These findings position $BaMnBi_2$ as a promising candidate for the further exploration of coupled spin-valley physics in bulk materials.

## II. EXPERIMENTAL DETAILS AND METHODS

The $BaMnBi_2$ crystals used in this study were grown using the Bi-flux method. The starting materials of Ba, Mn, and Bi metals were mixed in a small alumina crucible at a ratio of Ba:Mn:Bi = 1: 1: 4 or 1: 1: 5 and then sealed in an evacuated quartz tube. The ampoule was heated to 1000 °C and held at that temperature for 1 day, followed by a slow cooling process at a rate of 2 °C/h to 400 °C. The ampoule was inverted, and the flux and single crystals were separated by a centrifuge. Plate-like single crystals as large as a few millimeters were obtained. X-ray diffraction (XRD) was utilized to analyze and confirm the crystal structure of the synthesized samples. Supplementary Fig. 3 presents the XRD pattern measured on the (001) plane. The sharp (00*l*) peaks demonstrate good crystalline quality of the grown crystals.

$BaMnBi_2$ crystals exhibit reduced air stability compared to $BaMnSb_2$ crystals. Hence, the crystals were handled carefully to minimize degradation and ensure measurement reliability. Crystals were stored in an argon-filled glovebox, and samples were prepared and measured within one hour of exposure to atmosphere. For high-field measurements, GE varnish was applied over the samples to encapsulate the sample after electrical contacts were made using silver epoxy, which significantly improved the sample's air stability during measurements at the magnet lab.

High-field transport measurements were performed at the National High Magnetic Field Laboratory (NHMFL) DC-field facility in Tallahassee, FL, USA. The data presented in this paper for magnetoresistivity and Hall resistivity were obtained by symmetrizing and anti-symmetrizing the longitudinal and transverse Hall resistivity measurements taken at positive and negative magnetic fields, respectively. Nonlinear transport measurements were performed using a lock-in technique with a Quantum Design Physical

Property Measurement System (PPMS). A Keithley 6221 precision AC/DC current source provided the driving current. AC and DC voltages were measured using the Stanford Research SR860 lock-in amplifier.

## III. RESULTS AND DISCUSSION

### A. Quantum Hall effect

Figure 1c shows a representative crystal image captured with a polarized light microscope, clearly revealing strip-like domains. This observation indicates that the structure of our $BaMnBi_2$ crystals is orthorhombic, as recently determined by Kondo et al. [21], rather than the previously reported tetragonal structure [25]. From the domain wall orientation, the crystallographic *a* (or *b*) axis can be determined. Specifically, the *a*- (or *b*-) axis forms a 45° angle relative to the domain wall, and the *a*/*b*-axis is rotated by 90° across the domain wall, as illustrated in Fig. 1c. Due to the minimal difference between the lattice parameters of *a* and *b*, distinguishing them using a Laue pattern is impossible. Since the orthorhombic distortion in $BaMnBi_2$ is much smaller than in $BaMnSb_2$, it leads to a unique spin-valley locked state, distinct from that in $BaMnSb_2$, as illustrated in Fig. 1b. Given that our prior work demonstrated the efficacy of electric transport measurements in probing spin-valley degeneracy via the QHE [18] and valley-contrasted Berry curvature via the NLHE [26] in $BaMnSb_2$, we conducted similar transport studies on $BaMnBi_2$. Our aim was to reveal transport signatures of spin-valley locked Dirac fermions in $BaMnBi_2$ by investigating its potential QHE and NLHE.

We performed magnetoresistivity measurements on two representative samples of $BaMnBi_2$ with the electric current applied to the *a*/*b*-axis (Sample #A) and *c*-axis (Sample #C) respectively, under high magnetic fields of up to approximately 35T (see the insets to Fig. 1d-f for the experimental configurations). Figure 1d-f illustrates the magnetic field dependence of Hall resistivity ($\rho_{xy}$), in-plane resistivity ($\rho_{xx}$), and out-of-plane resistivity ($\rho_{zz}$) measured at a temperature of 1.4 K respectively. All these data exhibit SdH oscillations starting from around 5T. It is evident from Fig. 1d-e that when $\rho_{xx}$ reaches minima at about 10 and 23T, $\rho_{xy}$ displays distinct plateau features as marked by the horizontal/vertical dashed lines, providing an implication of bulk QHE in $BaMnBi_2$ (Note that the Hall resistivity data in Fig. 1d is presented with -$\rho_{xy}$ for better visualizing the plateau characteristics). Furthermore, we also note that the SdH oscillations of $\rho_{zz}$ are out-of-phase with those of $\rho_{xx}$. A similar phenomenon was also observed in $BaMnSb_2$ and is attributed to distinct transport mechanisms between the in-plane and out-of-plane directions. Since $BaMnSb_2$ possesses a quasi-two-dimensional (2D) electronic band structure, its interlayer transport occurs via the tunneling process, in contrast with in-plane band transport [18]. According to the Lifshitz-Kosevich (LK) theory [27,28], $\rho_{xx}$ exhibits a minimum when the density of state at the Fermi level $E_f$ [DOS($E_f$)] reaches a minimum. The minimal DOS($E_f$) leads to a minimal interlayer tunneling conductivity such that $\rho_{zz}$ exhibits a maximum at minimal $\rho_{xx}$. Given that $BaMnBi_2$ shares a similar layered structure with $BaMnSb_2$, the electronic band structure of $BaMnBi_2$ should also be quasi-2D, which was indeed demonstrated in prior work [25]. Therefore, the interlayer tunneling mechanism should also apply to $BaMnBi_2$ and can explain why its $\rho_{zz}$ and $\rho_{xx}$ exhibits out-of-phase SdH quantum oscillations. This is in line with the large resistivity ratio between the out-of-plane and in-plane direction probed in $BaMnBi_2$ ($\rho_{zz}/\rho_{xx} \sim 115 – 140$ at 1.4 K, refer to Supplementary Figure 2 [29]) and further implies that the weak interlayer coupling may lead to the presence of stacked QHE in $BaMnBi_2$ as seen in $BaMnSb_2$ [18]. On the other hand, we note that $\rho_{zz}$ exhibits metallic-like temperature dependence (Supplementary Figure 2). This behavior is not necessarily incompatible with a tunneling-dominated transport mechanism. In general, when the tunneling barrier is low and transmission is coherent, the tunneling conductance can remain sufficiently high such that the overall temperature dependence of the junction resistance appears metallic-like, as observed, for example, in c-axis junctions prepared by cleaving Ru-containing $Sr_2RuO_4$ single crystals [30]. The metallic-like *c*-axis behavior observed in both $BaMnSb_2$ and $BaMnBi_2$ may arise from a similar scenario.

To find further evidence of the possible QHE in $BaMnBi_2$, we conducted not only magnetic field sweep measurements of $\rho_{xy}$ (Fig. 2a), $\rho_{xx}$ (Fig. 2e) and $\rho_{zz}$ (Fig. 2i) under various field orientations at a fixed temperature (1.4 K), but also field sweep measurements of $\rho_{xy}$ (Fig. 2c), $\rho_{xx}$ (Fig. 2g), $\rho_{zz}$ (Fig. 2k) at various temperatures under the magnetic field perpendicular to the *ab*-plane. Figure 2a illustrates that the values of the $\rho_{xy}$ plateaus remain constant regardless of the field orientation angle $\theta$ (see the inset to Fig. 2a), as highlighted by the two horizontal dashed lines. This suggests that the Hall plateaus depend solely on the effective field, i.e. the field component perpendicular to the *ab*-plane ($B_{\perp} = Bcos(\theta)$). This feature is further illustrated in Fig. 2b which shows the angular dependence of $\rho_{xy}$ at 10 T and 23 T where the Hall plateaus were observed. As depicted, the $\rho_{xy}$ value at 10 T remains nearly constant as $\theta$ increases from 0 to 40°, but decreases significantly for $\theta$> 40 °. At 23 T, $\rho_{xy}$ exhibits a similar angular dependence, with a slight decrease from 0 to 40° attributed to the valley splitting, as explained below. These results contrast sharply with the expected angular dependence of Hall resistivity for normal conductors where $\rho_{xy} \propto B\cos\theta$ (indicated by the dashed lines in Fig. 2b). However, they can be well understood in terms of the QHE. The angle-independent $\rho_{xy}$ at 10 T and 23 T within the $0 \leq \theta < 40°$ range suggests that the Hall resistivity quantization remains unchanged with the field rotation within this angle range. The angular dependences of $\rho_{xx}(B)$ (Fig. 2e-f) and $\rho_{zz}(B)$ (Fig. 2i-j) are in line with this argument. Specifically, when $\rho_{xy}(B)$ exhibits a plateau (Fig. 2a), $\rho_{xx}(B)$ and $\rho_{zz}(B)$ consistently shows a minimum and a maximum, respectively, regardless of $\theta$ (Fig. 2e & 2i). Moreover, as shown in Fig. 2f & 2j, the $\rho_{xx}$ and $\rho_{zz}$ values at 10 T display minimal variation with $\theta$ in the $0 \leq \theta < 40°$ range where $\rho_{xy}$ remains constant, but show significant variations for $\theta > 40°$, where $\rho_{xy}$ decreases noticeably. These unusual angular dependences of $\rho_{xx}$ and $\rho_{zz}$ apparently do not follow the $cos^2(\theta)$ dependence expected for trivial metals (indicated by the dashed lines in Fig. 2f & 2j) but align with the quantum Hall scenario. The large variation of the $\rho_{xx}$ and $\rho_{zz}$ values at 23 T within the $0 \leq \theta < 40°$ range is associated with valley splitting, as will be explained below.

The indication of QHE in $BaMnBi_2$ was also revealed through temperature dependent measurements of $\rho_{xy}(B)$ (Fig. 2c), $\rho_{xx}(B)$ (Fig. 2g) and $\rho_{zz}(B)$ (Fig. 2k). In Fig 2c, the $\rho_{xy}$ plateaus exhibit remarkable temperature independence up to 20K, while the quantum oscillations persist up to 75K. This characteristic is clearly shown in Fig. 2d which plots the $\rho_{xy}$ values at 10 and 23 T as a function of temperature. Similar to $\rho_{xy}$, $\rho_{xx}$ and $\rho_{zz}$ also display temperature independence below 20K under the field of 10 T (Fig. 2h & 2l), but they become temperature dependent under the field of 23 T. In ideal 2D quantum Hall systems, as $\rho_{xy}$ is quantized, $\rho_{xx}$ reaches zero due to the dissipationless transport supported by the chiral edge state. However, in most practical quantum Hall systems, where bulk states have minor contributions to transport, $\rho_{xx}$ just reaches a minimum as $\rho_{xy}$ exhibits a quantized plateau. In such cases, both $\rho_{xx}$ and $\rho_{xy}$ exhibit temperature independence, attributed to the dominant nonlocal edge transport supported by chiral edge states, which are not subject to backscattering. Conversely, when transport involves substantial bulk contribution, $\rho_{xx}$ and $\rho_{xy}$ become markedly temperature dependent. In stacked quantum Hall systems like $BaMnSb_2$, the quantum Hall state is also manifested in the temperature dependence of $\rho_{zz}$: the temperature independence of $\rho_{xx}$ coincides with the temperature independence of $\rho_{zz}$, though they show out-of-phase SdH oscillations. Our observation of temperature independence of $\rho_{xx}$, $\rho_{xy}$ and $\rho_{zz}$ at 10 T are consistent with the quantum Hall state. While not conclusive on its own, this observation motivates the QHE interpretation, for which the key supporting evidence is presented in the following paragraphs. At 23 T, while $\rho_{xy}$ exhibits temperature independent plateau below 20 K (Fig. 2d), both $\rho_{xx}$ and $\rho_{zz}$ exhibit significant temperature dependence even below 20 K (Fig. 2h & 2l), resulting from valley splitting, as discussed below. The QHE signatures seen in Sample #A were also reproduced in another sample #C, (see Supplementary Note 1 and Supplementary Fig. 1)

A detailed analysis of the $\rho_{xy}$ data at 1.4 K (Fig. 3a) reveals robust evidence for the presence of stacked QHE in $BaMnBi_2$. As shown in the inset to Fig. 3a, we first extract the step size of $1/\rho_{xy}$ plateaus, denoted as $1/\rho^0_{xy}$, and plot the normalized quantity $\rho^0_{xy}/\rho_{xy}$ as a function of $B_F/B$, where $B_F$ is the SdH oscillation frequency. We estimate $B_F$ to be 13T using $\rho_{xx}$ data (Fig. 2g) considering the valleys at 8.4T and 23.7T for the temperature above 15K, where the valley splitting vanishes. Near the minima of $\rho_{xx}$, $\rho^0_{xy}/\rho_{xy}$ is quantized to half integer values ($\frac{1}{2}$, $\frac{3}{2}$) corresponding to the half-integer normalized filling factor ($\frac{1}{2}$, $\frac{3}{2}$) given by $B_F/B$ at the $\rho_{xx}$ minima. The half-integer filling factor corresponds to the non-trivial Berry phase accumulated in the cyclotron orbits. These distinct features, combined with the angular independence of $\rho_{xy}$, $\rho_{xx}$ and $\rho_{zz}$ at 10 T for $\theta < 40°$ (Fig. 2b, 2f & 2j) and their temperature independence below 20K for $B \perp ab$ (Fig. 2d, 2h & 2l), are indicative of stacked QHE of Dirac fermions, where the 2D Bi conducting layers between the $Ba\text{-}MnBi_4\text{-}Ba$ insulating slabs act as effective quantum Hall layers. Although previous magnetotransport studies on $BaMnBi_2$ revealed SdH oscillations and its Dirac Fermion transport properties [21,25], its QHE was not demonstrated. Given that the quantum oscillation frequency is proportional to the extremal cross-section area of the Fermi surface, our observed small oscillation frequency (13 T for Sample #A and 11.5 T for sample # B) in $BaMnBi_2$ suggests that its Dirac band crossing points are positioned very close to the Fermi level.

Similar to the QHE of $BaMnSb_2$ [20], the bulk QHE of $BaMnBi_2$ should also originate from the parallel transport of 2D Bi Dirac layers stacked along the c-axis. Hence, the spin valley degeneracy per Bi layer *s* can be estimated using the equation $\frac{1}{\rho_{xy}^0} = sZ^*(e^2/h)$, where $Z^*$ represents the number of layers per unit length [31]. Given that a unit cell of $BaMnBi_2$ comprises of two Bi conducting layers, we can simplify the equation to $\frac{1}{\rho_{xy}^0} = s(2/c)(e^2/h)$, where *c* is the lattice parameter along the out-of-plane direction. Our calculation yielded spin valley degeneracy *s* to be 3.7. It is important to note that experimentally determined *s* generally has finite uncertainty due to factors such as inhomogeneous transport caused by dead layers (i.e. those Bi layers not showing QHE), imperfect contacts and/or errors in measuring sample dimensions. With these considerations, we approximate the spin valley degeneracy in $BaMnBi_2$ to be 4.

Further, we examined the validity of the spin-valley degeneracy of 4 by comparing the carrier density estimated from the quantum oscillation frequency $B_F$ with the transport carrier density extracted from the Hall coefficient. According to Luttinger's theorem, the carrier density of a 2D system with a degeneracy of 4 can be expressed as $n_{2D} = 4eB_F/h$, where *e* is the elemental charge and $h$ is the Planck's constant. Since each unit cell in $BaMnBi_2$ contains two conducting Bi layers (Fig. 1a), the 3D carrier density can be expressed as $n_{SdH} = n_{2D}/(c/2)$. The estimated $n_{SdH}$ value for the sample used for $\rho_{xx}$ and $\rho_{xy}$ measurements (i.e. sample #A) is $0.92\times10^{19}$ cm$^{-3}$ which is in good agreement with the carrier density determined by the Hall coefficient ($n_{Hall} \sim 0.69\times10^{19}$ cm$^{-3}$). This consistency between $n_{SdH}$ and $n_{Hall}$ provides additional support for a degeneracy of 4 in $BaMnBi_2$. However, as noted earlier, previous band structure calculations predicted four and two spin-polarized valleys near the X and Y points, respectively, as well as four additional valleys near the midpoint of the ΓM line (Fig. 1b, right panel) in $BaMnBi_2$ [21], suggesting an expected $s$ value of 10. The discrepancy between theory and experiment is likely due to (i) inhomogeneous bulk transport and/or imperfect contact, as indicated above, or (ii) the actual spin-valley locked state differing from that predicted in Ref [21]. Considering the consistency of the carrier density measured by the SdH oscillations and Hall effect in our experiment, scenario (i) seems less likely. Our observed SdH frequency closely matches the 13 T value reported in Ref. [21], which was attributed to the $\alpha$ Fermi pockets near the Y point, assuming a Fermi level shift of –33 meV. The extracted degeneracy factor of 4 is consistent with the presence of the $\alpha$ pockets predicted by band structure calculations.

From the temperature dependence of SdH oscillations, we also estimated the effective mass of Dirac fermions in $BaMnBi_2$. According to the LK theory, the oscillatory part of resistivity can be expressed as: $\Delta\rho \propto R_T R_D \cos\left[2\pi\left(\frac{B_F}{B} - \frac{1}{2} + \beta + \delta\right)\right]$, with temperature damping factor, $R_T = \frac{\lambda T}{\sinh(\lambda T)}$ and Dingle damping factor, $R_D = e^{-\lambda T_D}$, where $\lambda = (2\pi^2 k_B m^*)/(\hbar e B)$. $2\pi\beta$ is the Berry phase and $T_D$ is the Dingle temperature. $\delta$ is a phase shift, $\delta = 0$ and $\pm 1/8$ for the 2D and 3D systems, respectively. Figure 3b presents the analysis of the quantum oscillation data where we performed a fast Fourier transform (FFT) of the second derivative of $\rho_{zz}$ (see inset) and normalized the FFT amplitude along the dashed line. The temperature dependence of the normalized FFT amplitude can be accurately fitted with $R_T$ to obtain the effective mass of $m^* = 0.027m_0$. Such a small effective mass is a characteristic feature of Dirac fermions. The $m^*$ in $BaMnBi_2$ is lower than that of $YbMnSb_2$ ($0.134m_0$) [32], $Sr_{1-y}Mn_{1-z}Sb_2$ ($0.14m_0$) [33] and comparable to those of $CaMnSb_2$ ($0.05$-$0.06m_0$) [34] and $BaMnSb_2$ ($0.052$-$0.058m_0$) [35].

As shown in Fig. 1e, an extra peak appears in $\rho_{xx}$ near 33T at 1.4K (marked by the arrow) which gradually diminishes as the temperature increases above 10 K (Fig. 2g). Although our field range (< 35 T) does not allow us to resolve the full double-minimum structure reported by Kondo et al. [20], the similarity in temperature evolution suggests that this feature could be related to valley splitting. In conventional electron systems, such splitting in $\rho_{xx}$ minima is often attributed to Zeeman-driven Landau level splitting. However, our c-axis conductivity $\sigma_{zz}$ measurements at various tilt angles (Supplementary Fig. S4) show no discernible angular dependence, making a large Zeeman contribution unlikely. If the feature were Zeeman-induced, the effective *g*-factor would need to be on the order of 10–20, large enough to produce a strong angular dependence in $\sigma_{zz}$ — which we do not observe. Given the structural similarity between $BaMnBi_2$ and $BaMnSb_2$, we suggest that the observed feature may arise instead from a second-order $\boldsymbol{q}$-quadratic term [20], similar to what has been observed in $BaMnSb_2$, which can lift the spin-valley degeneracy and produce valley splitting.

### B. Nonlinear Hall measurements

The spin-valley locking model discussed above for $BaMnBi_2$ is further supported by our nonlinear Hall effect (NLHE) measurements. NLHE refers to a phenomenon where a longitudinal AC current ($I^{\omega}$) can induce a second-order Hall voltage ($V_{\perp}^{2\omega}$) along the transverse direction under the time-reversal symmetry condition, with $V_{\perp}^{2\omega}$ scaling quadratically with $I^{\omega}$ [36]. NLHE can have either intrinsic or extrinsic origins. The intrinsic origin is tied to the quantum geometry of the electron wave function. Theoretical predictions suggest that intrinsic NLHE occurs when electronic bands exhibit a net Berry curvature dipole (BCD) [37] or quantum metric [38,39], which correspond to the imaginary and real parts of the quantum geometric tensor, respectively. BCD-induced NLHE is typically expected in non-centrosymmetric materials with tilted Dirac/Weyl cones [37], while quantum metric-induced NLHE appears in magnetic materials that break both inversion symmetry (*P*) and time-reversal symmetry (*T*), with or without *PT* symmetry [38,39]. BCD-induced NLHE has been observed in various 2D systems, including few-layer $WTe_2$ [36,40], engineered bilayer graphene [41], and strained or twisted $WSe_2$ monolayers [42,43]. In 3D materials, related responses have been observed in $BaMnSb_2$ [26], Dirac/Weyl semimetals [44], and polar Weyl systems such as $TaIrTe_4$ [45]. Extrinsic NLHE can result from skew scattering and/or side-jump mechanisms in non-centrosymmetric materials [46,47], or from geometric asymmetric scattering due to textured nanoparticles [48].

Similar to $BaMnSb_2$ [26], $BaMnBi_2$ is expected to exhibit valley-contrasting Berry curvature due to its spin-valley locked Dirac cones, leading to a net BCD. Therefore, an intrinsic NLHE is expected for $BaMnBi_2$. To test this, we performed nonlinear transport measurements using lock-in techniques. Our results showed clear second-harmonic Hall voltage ($V_{\perp}^{2\omega}$) responses driven by alternating currents (Fig. 4a). The measured $V_{\perp}^{2\omega}$ displayed no frequency dependence up to 1800 Hz (Supplementary Fig. 5) and scaled quadratically with the applied current $I^{\omega}$ in the current range below 5 mA at 2 K (Fig. 4b), which is consistent with NLHE behavior. Furthermore, as shown in Fig. 4a, the second-harmonic Hall voltage reversed its sign when both the current direction and the Hall probe were simultaneously reversed, aligning with the expected characteristics of NLHE [40].

Another hallmark of BCD-induced NLHE is the nonlinear angle $\theta_{NLHE}$ of 90°, $\theta_{NLHE}$ is defined as arctan($V_{\perp}^{2\omega}/V_{//}^{2\omega}$), where $V_{\perp}^{2\omega}$ represents the nonlinear Hall voltage and $V_{//}^{2\omega}$ the longitudinal voltage. As $V_{//}^{2\omega}$ = 0, $\theta_{NLHE}$ = 90°; that is, the second order response is present only along the transverse direction. Based on the $V_{\perp}^{2\omega}$ and $V_{//}^{2\omega}$ data presented in Fig. 4a, $\theta_{NLHE}$ is estimated to be 74°, only 16° deviation from the value of 90° in the ideal geometry. This deviation can be attributed to a small longitudinal second-harmonic signal caused by geometric mixing, as seen in Fig. 4(a), caused by the finite device aspect ratio and unavoidable misalignment between voltage probes and crystallographic axes. This type of mixing is often seen in nonlinear transport measurements and has also been reported in other materials [45]. The second-harmonic Hall voltage increases steadily as the temperature decreases (see Supplementary Fig. 9a), reaching 0.6 μV at 2K. Notably, the second-harmonic signal remains detectable even at room temperature (300K) (Supplementary Fig. 10), though it is substantially smaller. A stronger room-temperature NLHE might be achieved by optimizing the sample domain and increasing the current density, as discussed later. The quadratic current-voltage characteristic and its frequency independence, as well as the near-90° nonlinear angle indicate that our observed 2nd order response in $BaMnBi_2$ is not likely due to extrinsic effects such as junction or thermoelectric effects but has intrinsic BCD contribution (see Supplementary Note 3 for detailed discussions). To further substantiate the intrinsic origin, we not only reproduced the NLHE phenomenon in another sample (see Supplementary Fig. 6), but also found the scaling analysis of the nonlinear Hall conductivity ($\sigma_{yxx}^{(2)}$) is aligned with the expected behavior of BCD induced nonlinear

Hall effect, i.e. $\sigma_{yxx}^{(2)}$ is approximately proportional to longitudinal conductivity $\sigma_{xx}$ (see Supplementary Note 3).

Both $BaMnSb_2$ and $BaMnBi_2$ exhibit antiferromagnetic order due to Mn moments. Prior transport and band-structure studies indicate, however, that the Mn magnetism only weakly perturbs the low-energy Dirac bands: neither the Fermi surface nor the resistivity show a clear anomaly at the Néel temperature $T_N$ ≈ 288K. In $BaMnX_2$ (X = Bi, Sb), the spin–valley-coupled Dirac states near $E_F$ are primarily controlled by spin–orbit coupling on the Bi/Sb $p$-orbitals [49] and the polar lattice distortion [21], rather than by a strong exchange field from the Mn sublattice. For our analysis of the NLHE, we therefore neglect the small influence of the AFM order on the Dirac bands and treat $BaMnBi_2$ as effectively time-reversal symmetric. This treatment is consistent with our measurements; the nonlinear Hall response shows no anomaly across $T_N$ and the NLHE varies smoothly through $T_N$, confirming that the magnetic order does not play a role in the Berry-curvature–related physics probed in our study. $BaMnSb_2$ also shows a similar behavior, as discussed in our prior work [26].

While the second harmonic Hall signal at 10K follows the expected quadratic behavior characteristic of NLHE, the signal at 2 K begins to deviate from this quadratic trend above 5mA as noted above (Fig. 4b). This cannot be attributable to the heating effect in the sample, since the first harmonic signal measurements (inset to Fig. 4b) are linear; the resistance remains unchanged up to 10mA input current. The data is well fitted by incorporating a fourth-order correction, as illustrated by the black fitted curve in Fig 4b. This fourth order term can be associated with high-order nonlinear response [50,51]. Theory demonstrates that the transverse electric field from the nonlinear Hall effect influences the longitudinal conductivity through a bulk electric field effect, that could result in additional fourth order term in transverse direction with respect to applied current [50].

It is important to highlight that the NLHE of $BaMnBi_2$ is significantly suppressed by the presence of 90° and 180° domains. Note that the sample used in our current measurements has dimensions of 1-2 mm which includes numerous domains (see the image of our device in supplementary Fig. 8). According to theory [37], for a polar material, maximum nonlinear signal is generated when a current is applied perpendicular to the in-plane polar axis, while no signal is observed when the current is parallel to the polar axis. This means that 90° domains contribute no nonlinear signal, while 180° domains cancel out the overall NLHE signal. Consequently, the NLHE in $BaMnBi_2$ is diminished, as we observe numerous twin domains through a polarized microscope, with lengths on the order of 10 μm (Fig. 1c). The domain sizes are small in $BaMnBi_2$ because the orthorhombicity of the crystal structure is small. Finally, it should be noted that the second order nonlinear response is quadratically dependent on the current density. Consequently, samples with smaller cross section areas are more favorable for observing a significant nonlinear signal. In our measurements, we used devices with thickness of around 100 μm which explains the need for high current to extract observable nonlinear signals. For future studies, microscale or thin film devices on a single domain in $BaMnBi_2$ would be anticipated to significantly enhance the nonlinear signal by several orders of magnitude. This approach holds promises for advancing the exploration of nonlinear phenomena and potential applications, such as terahertz detection and energy harvesting, in $BaMnBi_2$-based devices.

## IV. CONCLUSION

In summary, we observed both bulk QHE and NLHE in the layered Dirac material $BaMnBi_2$. Our analysis of the QHE revealed a spin-valley degeneracy of 4. Additionally, we identified an intrinsic contribution to the NLHE, indicating that the Dirac bands exhibit a Berry curvature dipole. This finding aligns with theoretical predictions that $BaMnBi_2$'s electronic structure features valley-dependent spin polarization [21]. This spin-valley locked state contrasts with that of $BaMnSb_2$, which features spin-polarized valleys near the X point and a degeneracy of 2. These results underscore $BaMnBi_2$ as a distinctive candidate for exploring spin-valley coupled physics in bulk materials and for potential electronic device applications that leverage both spin and valley degrees of freedom with topological protection.

**ACKNOWLEDGEMENTS**

This work is supported by the U.S. National Science Foundation (NSF) under award No. DMR-2211327. L. Min and Z.Q.M. also acknowledge the partial support from NSF through the Materials Research Science and Engineering Center DMR 2011839 (2020 - 2026). The work at the National High Magnetic Field Laboratory is supported by the NSF Cooperative Agreement No. DMR1157490 and the State of Florida. Y.D.G. acknowledges the support by the US Department of Energy under grants DE-SC0019068 for performing high-field transport measurements at the National High Magnetic Field Lab.

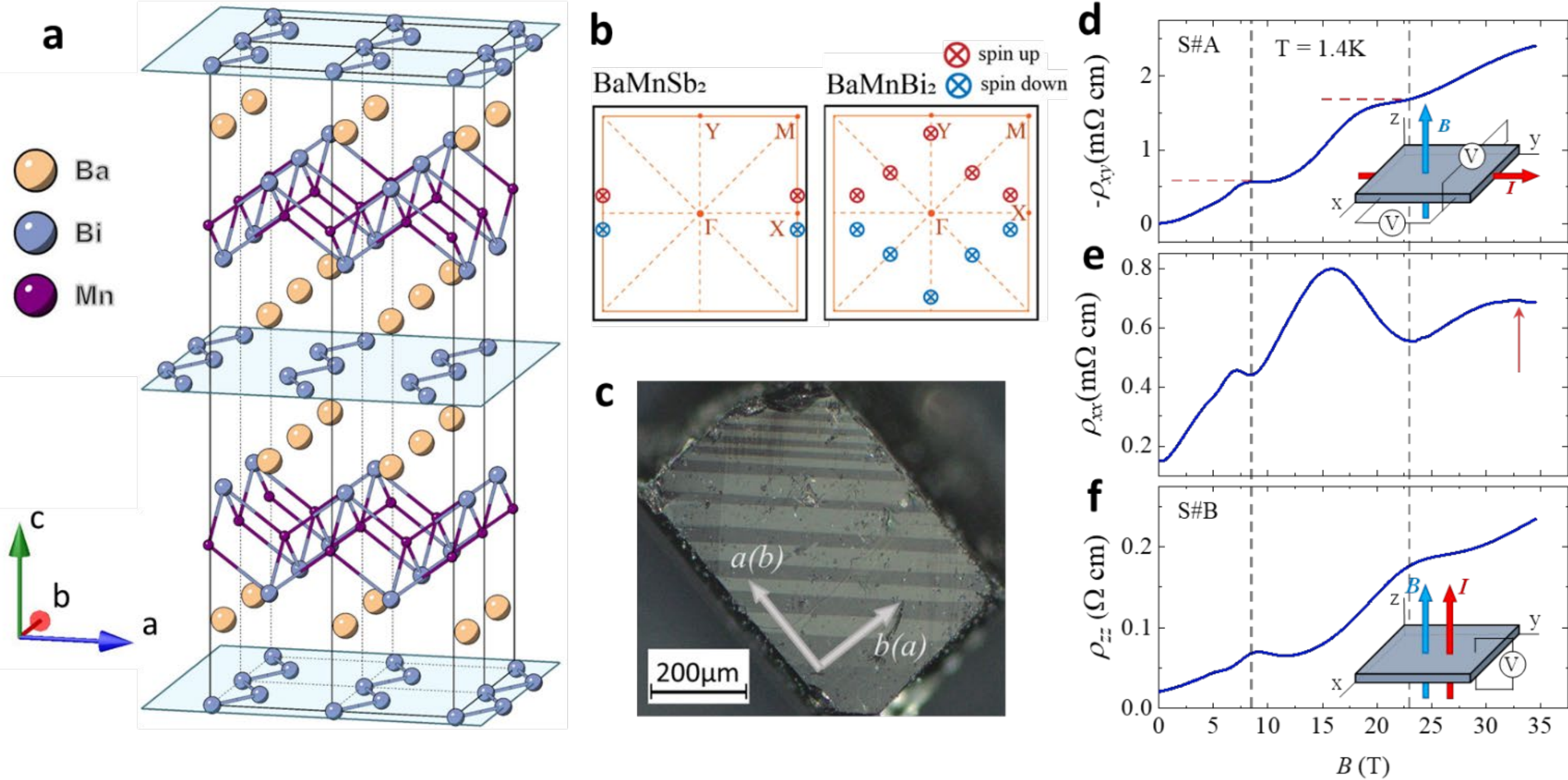

**FIG 1. Crystal structure, electronic structure and magnetoresistivity of $BaMnBi_2$. (a)** Crystal structure for $BaMnBi_2$. **(b)** Schematics of Dirac crossing points situated in the Brillouin zone of $BaMnBi_2$ and $BaMnSb_2$, as previously reported [18,20,21]. **(c)** The polarized microscope image of the as-grown (001) crystal surface. **(d, e, f)** Hall resistivity ($\rho_{xy}$), in-plane resistivity ($\rho_{xx}$), out-of-plane resistivity ($\rho_{zz}$) vs. magnetic field $B$ up to 35T at 1.4 K. The inset in (**d**): schematic of the experimental set-up for the $\rho_{xy}$ and $\rho_{xx}$ measurements on Sample #A (S#A). The inset in (**f**): schematic sample configuration for the $\rho_{zz}$ measurements on Sample #B (S#B). The dashed lines correspond to quantum Hall states. The arrow in (e) shows the local maximum associated with valley splitting.

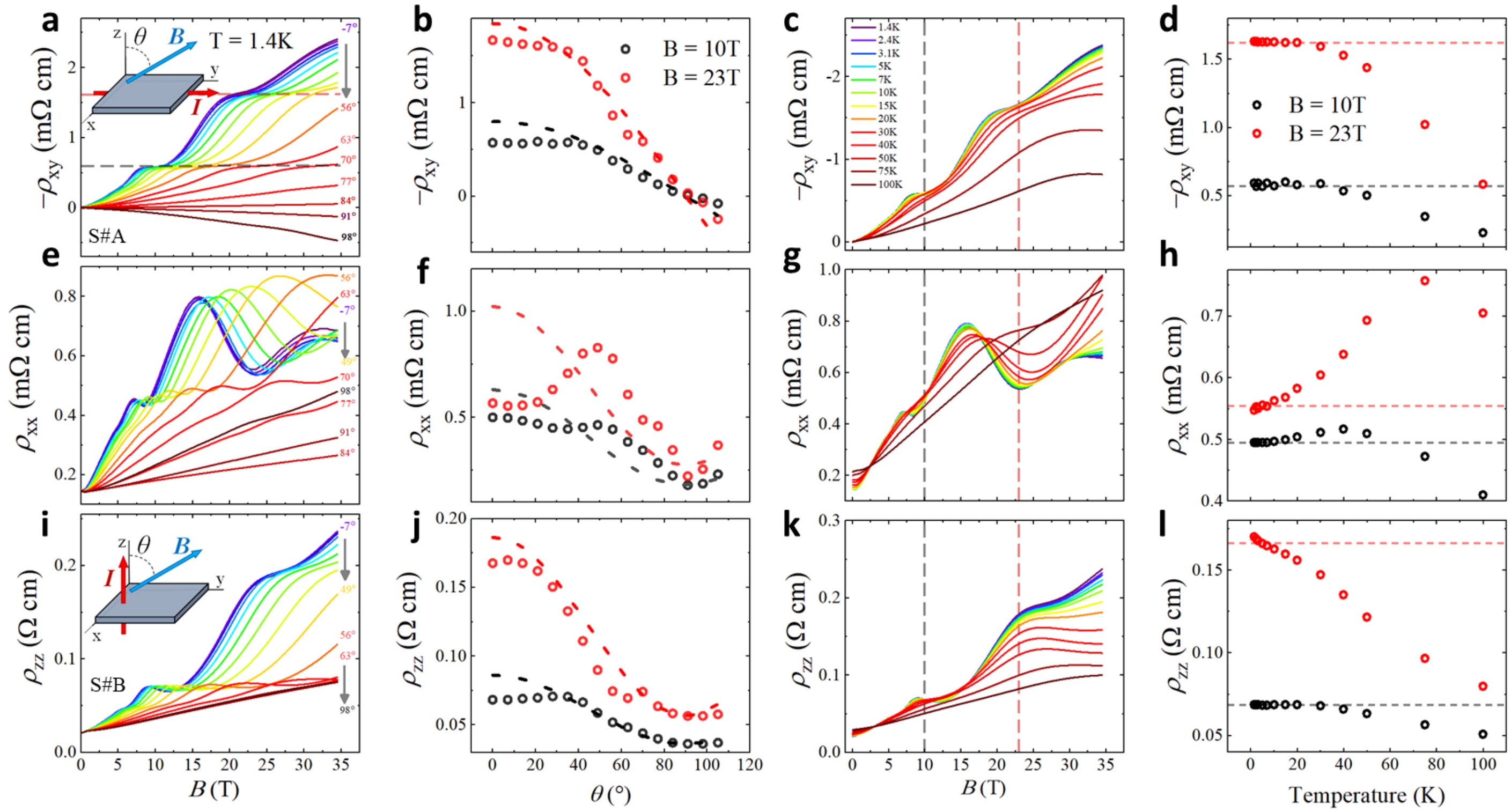

**FIG 2. Magnetoresistiviy at high fields with different field orientation and temperature. (a, e, i)** Angular ($\theta$) dependence of $\rho_{xy}$, $\rho_{xx}$ and $\rho_{zz}$ vs field $B$ at 1.4 K. Inset in (**a & i**): Schematic sample configuration for the $\rho_{xy}$, $\rho_{xx}$ and $\rho_{zz}$ measurements on S#A and S#B. **(b, f, j)** Angular dependence of $\rho_{xy}$, $\rho_{xx}$ and $\rho_{zz}$ at $B$ = 10T and 23T. The dashed lines represent the behavior expected for trivial metals. **(c, g, k)** Magnetic field dependence of $\rho_{xy}$, $\rho_{xx}$ and $\rho_{zz}$ at various temperatures. Dashed lines correspond to the quantum Hall states at $B$ = 10T and 23T. **(d, h, l)** Temperature dependence of $\rho_{xy}$, $\rho_{xx}$ and $\rho_{zz}$ at $B$ = 10T and 23T.

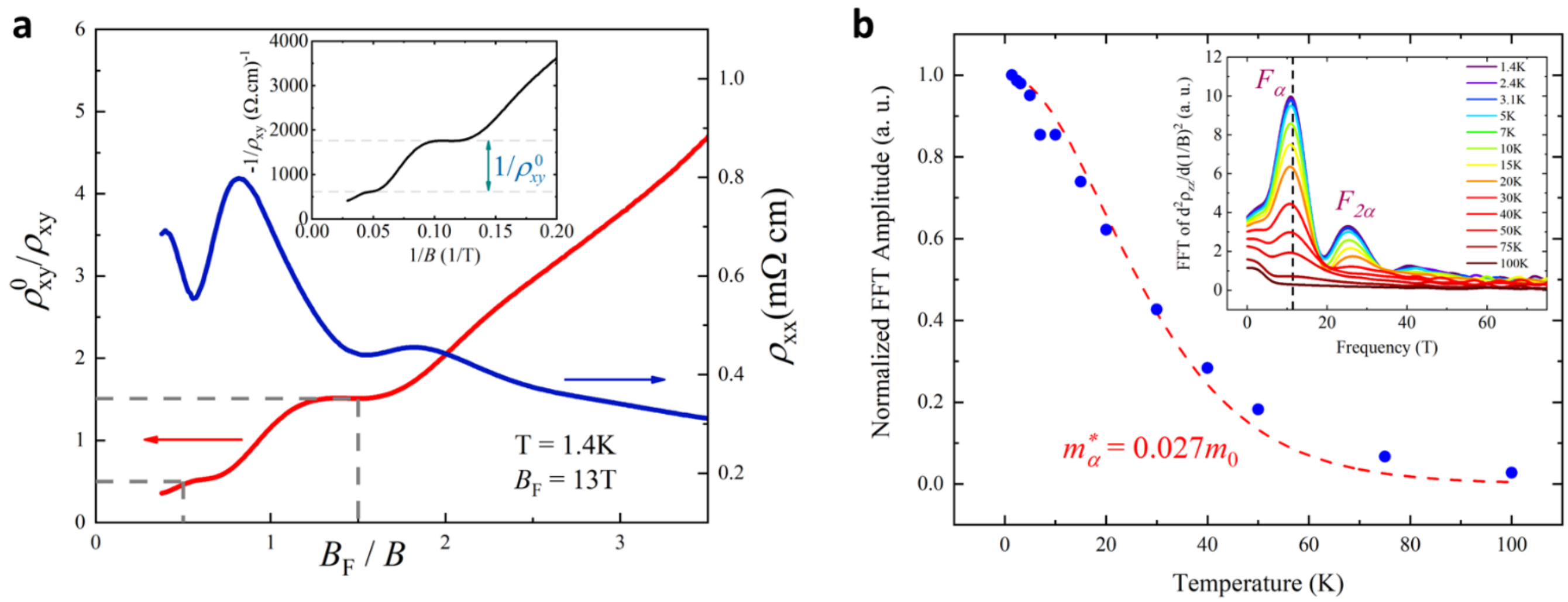

**FIG 3. Bulk quantum Hall effect and effective mass analysis based on SdH oscillations. (a)** Normalized inverse Hall resistivity ($\rho^0_{xy}$ / $\rho_{xy}$) and in-plane resistivity ($\rho_{xx}$) vs $B_F/B$ at 1.4K. The inset shows the -1/$\rho_{xy}$ as a function of 1/$B$. 1/$\rho^0_{xy}$ is determined by the vertical interval between steps. **(b)** Temperature dependence of normalized FFT amplitudes fitted to the LK formula. The inset shows the spectra of the FFT of second derivatives of $\rho_{zz}$ with respect to 1/$B$ at various temperatures for S#B. The dashed line represents the first harmonic peak at which the FFT was normalized for the fitting.

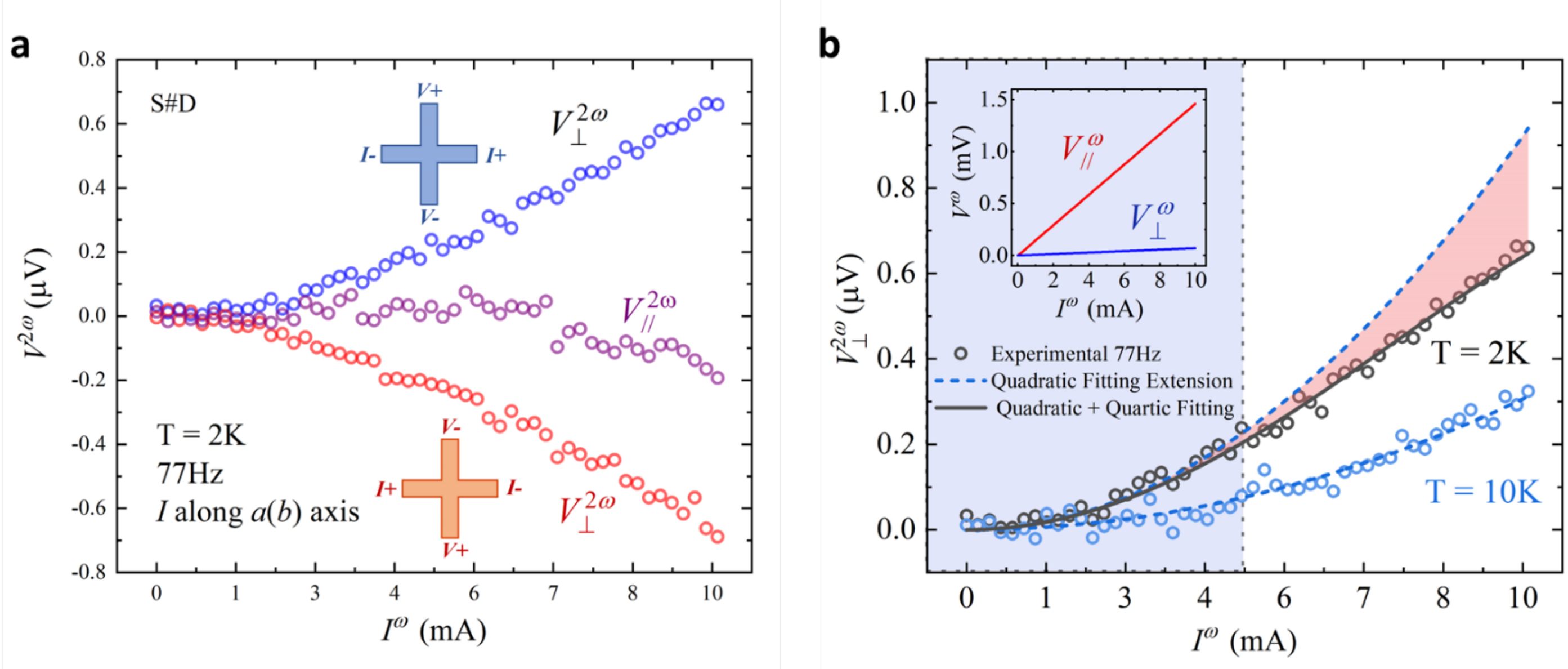

**FIG 4. Nonlinear Hall effect in $BaMnBi_2$. (a)** The second harmonic Hall voltage $V_\perp^{2\omega}$ vs the input AC current at 2 K. $V_{//}^{2\omega}$ represents the second harmonic longitudinal voltage. The schematics show the current and voltage leads configurations. When the current and voltage leads are reversed simultaneously, the second-harmonic Hall voltage changes sign. **(b)** $V_\perp^{2\omega}$ vs $I^\omega$at 77.77Hz frequency at 2 K and 10 K. The dotted curve shows the quadratic fits, and the solid black curve shows quadratic + quartic fit. The blue shaded region shows the region where quadratic curve well fits the $V_\perp^{2\omega}$ data for 2K. The inset shows first harmonic longitudinal and transverse voltage vs the input current at 2 K.